# The Magnetic Structure of Saturn's Magnetosheath

Ali H. Sulaiman,[1] Adam Masters,[2] Michele K. Dougherty,[1] and Xianzhe Jia[3]

Corresponding author: A.H. Sulaiman, Space and Atmospheric Physics, Blackett Laboratory, Imperial College London, London, UK. (ali.sulaiman08@imperial.ac.uk)

[1]Space and Atmospheric Physics, Blackett Laboratory, Imperial College London, London, UK.

[2]Institute of Space and Astronautical Science, Japan Aerospace Exploration Agency, 3-1-1 Yoshinodai, Chuo-ku, Sagamihara, Kanagawa 252-5210, Japan.

[3]Department of Atmospheric, Oceanic and Space Sciences, University of Michigan, Ann Arbor, Michigan, USA.

**Abstract**

A planet's magnetosheath extends from downstream of its bow shock up to the magnetopause where the solar wind flow is deflected around the magnetosphere and the solar wind embedded magnetic field lines are draped. This makes the region an important site for plasma turbulence, instabilities, reconnection and plasma depletion layers. A relatively high Alfvén Mach number solar wind and a polar-flattened magnetosphere make the magnetosheath of Saturn both physically and geometrically distinct from the Earth's. The polar flattening is predicted to affect the magnetosheath magnetic field structure and thus the solar wind-magnetosphere interaction. Here we investigate the magnetic field in the magnetosheath with the expectation that polar flattening is manifested in the overall draping pattern. We compare an accumulation of Cassini data between 2004 and 2010 with global magnetohydrodynamic (MHD) simulations and an analytical model representative of a draped field between axisymmetric boundaries. The draping patterns measured are well captured and in broad agreement for given upstream conditions with those of the MHD simulations (which include polar flattening). The deviations from the analytical model, based on no polar flattening, suggest that non-axisymmetry is invariably a key feature of the magnetosphere's global structure. Our results show a comprehensive overview of the configuration of the magnetic field in a non-axisymmetric magnetosheath as revealed by Cassini. We anticipate our assessment to provide an insight to this barely studied interface between a high Alfvénic bow shock and a dynamic magnetosphere.

## 1. Introduction

The magnetosheath of a planet is the region between the interplanetary magnetic field (IMF) and the planetary magnetosphere. It is bounded by the bow shock, which deflects and slows down the solar wind, and the magnetopause obstacle, which the flow diverts around and the convected magnetic field lines overall drape tangentially to. The region is therefore an important site for both plasma micro- and macro-processes such as turbulence, instabilities, magnetic reconnection and plasma depletion layers (PDL - close packing of magnetic field lines near the magnetopause surface) [*Lepping et al.*, 1981; *Russell*, 1976; *Violante et al.*, 1995]. Micro-processes such as reconnection occur when the diffusive term becomes dominant over the convective term of the magnetic Reynolds number i.e. $\eta > UL$ (where $\eta$ is the magnetic diffusivity, $U$ the flow velocity and $L$ the characteristic length of the plasma structure). This breaks down the frozen-in flux condition which states that the magnetic field lines are 'frozen' into and move with the plasma fluid.

The bow shock is formed as a result of the solar wind flowing at a greater relative speed to the obstacle than the speed at which information about the obstacle's presence can be propagated through the fluid i.e. the fast magnetosonic speed. The supersonic solar wind is rapidly slowed down and heated into a subsonic regime immediately downstream of the shock, with the effect greatest at the subsolar region where the shock front is normal to the flow. Analogous to the diverging segment of a 'de Laval' nozzle, the magnetosheath cross-sectional area increases with solar zenith angle. The flow is then further driven by the associated tension in the draped IMF in addition to the substantially lower back pressure at the terminator region far downstream. This accelerates the flow as it diverts around the magnetosphere and continues to do so until it is supersonic and the 'freestream' solar wind conditions are eventually met.

Studies of the magnetosheath merge important subject areas of both collisionless shocks and magnetospheric physics. Its field and particle conditions are both an end result, to characterize the nature of a bow shock, and a prerequisite, to understand magnetospheric dynamics via mass, energy and momentum transfers from the solar wind to planetary magnetospheres. Draping of magnetic field lines is one of the processes canonically understood to influence magnetic reconnection, and thus also the extent of the magnetosheath plasma depletion layer [*Dungey*, 1961; *Sonnerup*, 1974; *Zwan and Wolf*, 1976]. This is a condition widely accepted for the terrestrial magnetopause. Further studies corroborate this importance on reconnection onset at other planets such as at Mercury [*Slavin et al., 2009*] and more recently *DiBraccio et al.* [2013] showed that reconnection occurs for a wide range of magnetic shear angles, the rotation of the magnetic field from the magnetosheath into the magnetosphere, likely because of the low $\beta$ conditions. *Masters et al.* [2014] on the other hand show no PDL response to cross-magnetopause magnetic shear because the magnetic flux transport rates associated with reconnection are too low to have any effect. Nonetheless, it is expected that the IMF orientation strongly controls where reconnection is occurring because of the $\beta$-imposed constraint of close to anti-parallel fields required for reconnection onset [*Masters et al.*, 2012].

One of the earliest studies of the magnetosheath predicted the draping of the magnetic field using a gas-dynamic model [*Spreiter et al.*, 1966; *Spreiter and Stahara*, 1980]. This was achieved using the hydrodynamics of a single-fluid, nondissipative gas to describe the bulk flow around a planet with a prescribed non-self-consistent magnetic field convecting in unison. Further work was carried out by *Fairfield* [1969] and *Crooker et al.* [1985] where they compared this model to Earth observations of draping in the dayside magnetosheath. The latter used a multi-spacecraft technique with ISEE 3 upstream measuring the IMF as inputs to the gas-dynamic model and compared with time-lagged observations by ISEE 1 in the magnetosheath. They concluded that the observed draping near the dayside magnetopause is relatively consistent with the simple gas-dynamic model.

The magnetosheath of Saturn has commonalities to that of the Earth and Jupiter but also a significant uniqueness [*Richardson*, 2002; *Sergis et al.*, 2013]. It is distinctive in its (dayside) geometry which is governed by the competing anti-planetward pressures, due to internal magnetospheric processes [*Achilleos et al.*, 2008], impinging on the magnetopause and the dynamic pressure of the much more tenuous solar wind plasma upstream of the bow shock. Two main features of its global geometry are the subsolar thickness, dictated by both internal and external pressure variability, and non-axisymmetry of the magnetopause by polar flattening; both of which are expected to have an effect on the magnetosheath's structure. An analytical treatment has been developed using ideal MHD to describe the conditions in the magnetosheath of a non-axisymmetric magnetopause [*Erkaev et al.*, 1996; *Farrugia et al.*, 1998]. The IMF orientation was found to play an important role in controlling the properties of the magnetosheath as a consequence of the deviation from axisymmetry. The magnetic field in the magnetosheath was shown to compress with the effect most pronounced nearer the magnetopause. The field lines were shown to rotate towards the planet's rotation axis, an additional effect only present in a non-axisymmetric case. This was also found to have a significant influence on the plasma flow, with the acceleration triggered by the magnetic tension force in a direction perpendicular to the magnetic field lines [*Erkaev et al.*, 2011, 2012]. Saturn's magnetosheath is also physically different with a high plasma $\beta$ (ratio of particle to magnetic field pressures) environment [*Masters et al.*, 2012] owing to the high Alfvén

Mach number bow shock. Both the Mach number and magnetosheath plasma $\beta$ increase monotonically with heliocentric distance.

In this paper, we investigate the large-scale overall configuration of Saturn's magnetosheath magnetic field using observations made by the Cassini spacecraft. While ongoing studies of high-latitude Cassini orbits aim to constrain the extent of magnetospheric polar flattening, here we present the magnetic field structure of the magnetosheath which is largely at lower latitudes. We compare and contrast the magnetic field observations with outputs from the BATSRUS MHD model in each of the equatorial and meridional planes and further compare four cases when the IMF orientation was relatively steady while Cassini traversed the magnetosheath with an analytical model describing draping between axisymmetric boundaries.

## 2. Cassini Observations

**2.1 Data Selection**

In this section, we introduce the type of data selected, the applicability and limitations of models used, and the approaches to conduct the analyses. We use data obtained from Cassini's onboard fluxgate magnetometer (MAG) [*Dougherty et al.*, 2004] from which boundary crossings and magnetosheath signatures are identified. Since we are interested in the large-scale spatially-dependent structure of the magnetosheath, we have selected 106 complete and uninterrupted magnetosheath traversals from Saturn Orbit Insertion (SOI) in 2004 to 2010 inclusive. These are both inbound (bow shock to magnetopause) and outbound (magnetopause to bow shock) and exclude excursions due to global boundary oscillations or surface waves. Such excursions are generally identified as a series of crossing over a timescale much shorter than the magnetosheath traversal [*Mistry et al.*, 2014].

The coordinate system used throughout this study is the Cartesian Kronocentric Solar Magnetospheric (KSM) system which centres Saturn at the origin, with positive *X* pointing towards the Sun, *Y* orthogonal to the magnetic dipole axis (approximately aligned with the rotation axis at Saturn) and pointing towards dusk, and *Z* chosen such that the magnetic dipole axis is contained in the *X-Z* plane with positive *Z* pointing north. A Saturn radius is the unit of distance ($R_S$; 1 $R_S$ = 60,268 km). Figure 1 highlights the spacecraft positions relative to Saturn where the magnetosheath is observed. The total coverage sums up to 2,486 hours with 84% of this on the dayside, 65% and 35% on the dawn (<1200 LT) and dusk (>1200 LT) flanks respectively, and particularly limited to lower and equatorial latitudes. The magnetic field and position measurements used are at 1-min resolution. This is sufficient and a higher time resolution does not improve the analysis since, given the range of time over which the analysis is conducted, the adjacent samples are not likely to be statistically independent. The angles are measured as the meridional angle which has a range of -90° (southward) to +90° (northward) and the azimuthal angle which is calculated counter-clockwise from +*X* and has a range 0° to 360° (see insets in Figure 1).

**2.2 Upstream Conditions and the Overall Configuration of the Magnetic Field in Saturn's Magnetosheath**

Cassini is presently the only spacecraft probing the kronian vicinity. This poses a difficulty in magnetosheath analyses since in-situ measurements of upstream conditions, which play an important role in driving the magnetosheath structure and dynamics, cannot be obtained simultaneously. In most cases, the timescales of solar wind variability are small compared to Cassini's time of flight through the magnetosheath, meaning the spacecraft is measuring particle and field parameters influenced by upstream conditions different to those measured just before the spacecraft crossed the bow shock (say for an inbound flight path). The bow shock and magnetopause boundaries exhibit global oscillations principally driven by the variability in the solar wind dynamic pressure resulting in changes to the location of the magnetosheath relative to Saturn (further away at lower dynamic pressures and vice versa). In addition, the two boundaries may respond differently to the solar wind dynamic pressure [*Slavin et al.*, 1985; *Hendricks et al.*, 2005] hence a change in the dynamic pressure is not necessarily proportional to the magnetosheath subsolar and polar thicknesses i.e. the planetocentric distances between the magnetopause and the bow shock at $X_{KSM} = 0$ and $Z_{KSM} = 0$ respectively (neglecting aberration).

Cassini sampled the upstream environment of Saturn for over a year before Saturn Orbit Insertion and the IMF orientation was measured to exhibit a bimodal distribution of the Parker spiral (in the ecliptic plane) angle averaging 90.6° ± 0.4° and 276.4° ± 0.5° (for angles < 180° and ≥ 180° respectively) with a slight meridional (out of the ecliptic plane) angle averaging 1.4° ± 0.3° [*Jackman et al.*, 2008].

Figure 2 illustrates the distribution of the observed configuration of the magnetic field in the magnetosheath for both the dawn and dusk flanks. The bin size for all plots is 10°. The color plots are 2D histograms consolidating both meridional and azimuthal angles. The color bar represents the length of time the angles were observed to be in a particular orientation. The histograms on their sides project the distributions of the individual angles. The orientation in the dawn flank shows a bimodal distribution of the azimuthal angle, with peaks shifted to the left from the Parker spiral, averaging at 65.9° ± 0.2° and 245.5° ± 0.1° (for angles < 180° and ≥ 180° respectively). The dusk flank has peaks shifted to the right from the Parker spiral and averaging at 112.9° ± 0.2° and 297.4° ± 0.2°. The relative amplitude of each pair of peaks indicate the ratio of time Cassini spent on the inward and outward pointing regions either side of the Heliospheric Current

Sheet. The meridional angles average at 1.9° ± 0.1° for the dawn flank and a more substantial 12.1° ± 0.2° for the dusk flank. The distributions are also more dispersed than that of the IMF upstream of Saturn [*Jackman et al., 2008*].

## 3. Results

### 3.1 MHD Simulations

We compare the magnetic field observations to the global BATSRUS MHD model which solves the governing MHD equations using a conservative finite-volume method [*Gombosi et al.*, 2002; *Jia et al.*, 2012]. The model has been tailored to simulate the kronian environment and outputs are generated for two IMF limiting cases: duskward and northward. This will allow us to estimate the (predominantly dayside) angular change of the magnetic field with respect to longitude on the equatorial *X-Y* plane (for duskward IMF) and the angular change with respect to latitude on the meridional *X-Z* plane (for northward IMF). We do not place particular emphasis on the directions of the IMF, but rather on their alignments with the planes such that the third components orthogonal to the two planes are zero. These results are compared against observations to assess how well the MHD simulations capture and thus predict field line draping and to reveal any asymmetry between the two planes. The expectation is that polar-flattening of the magnetosphere is manifested on the draping pattern of the field lines and hence that the asymmetry can therefore be estimated.

Figure 3 is a snapshot of the two MHD simulated IMF configurations of duskward and northward viewed at $Z_{KSM} = 0$ and $Y_{KSM} = 0$ respectively where the magnetic field vectors are perfectly aligned with the planes. The precise location and shape of the magnetopause are ambiguous but the draping pattern of the field lines and hence their angular change with longitude and latitude can be clearly deduced. The upstream conditions of both runs are given in Table 1 and are set such that the IMF is initially northward and duskward respectively.

Figure 4 shows the distribution of the observed azimuthal angles against local time. There are two clear linear correlations showing organised draping in the equatorial plane. Near to the subsolar point (~1200 LT), the corresponding angles are ~90° (duskward) and ~270° (dawnward), indicative of the configuration at which the IMF is incident on the Saturnian magnetosphere and consistent with the expected directions of the Parker spiral. The gradient in the meridional angle with respect to local time for both duskward and dawnward IMF orientations are very similar and there is no indication of dependence on direction on a global scale. With increasing (decreasing) local time, the azimuthal angle of the magnetic field lines increases (decreases), asymptotically approaching the planar geometry of the magnetopause. The red line is the MHD simulated draping of a magnetic field line for a duskward configuration in the magnetosheath (see Figure 3a) taken at $Z_{KSM} = 0$ $R_S$. The angles deduced are averages at different local times in the magnetosheath proper. The MHD model is in good agreement with the observations and reveals a gradient of 0.47 degrees of azimuthal angle per degree longitude for $Z_{KSM} = 0$ $R_S$.

Quantifying the angular change per degree latitude on magnetic field lines in the meridional plane is more difficult since the range of latitudinal coverage is far more restricted than that of the longitudinal coverage. In addition, the IMF is statistically most likely to be incident at the Parker spiral angle which lies in the *X-Y* plane making it impossible to assess the latitudinal draping as there is no significant component in that plane. Figure 5a shows the distribution of the observed meridional angles against latitude. The observations are limited to lower latitudes of ~-10° to ~+20° and show a very high spread with no apparent organization such as in the equatorial plane. Therefore the latitudinal change of the meridional angle of the draped IMF, which is (quasi-) equatorial at most times, cannot be traced. Plotted in green are the means of five uncommon cases where the observed IMF are near (±45°) northward and relatively steady, with no abrupt and significant changes in both the direction and magnitude. The red plot is the MHD simulated draping of the magnetic field lines for a northward configuration (see Figure 3b) taken at $Y_{KSM}$ = 0 $R_S$. Near the subsolar point (~0° latitude), the corresponding meridional angle is ~90°, indicative of a northward IMF. There is some, albeit not conclusive, proximity of these five observed events to the MHD simulated plot, however these statistics are notably limited and consist of a high spread. The gradient from the MHD plot reveals a draping of 0.54 degrees per degree latitude for $Y_{KSM}$ = 0 $R_S$.

**3.2 Correspondence between observations and predictive model**

*Kobel and Flückiger* [1994] developed an analytical model (herein referred to as KF94) to characterize the magnetic field in the magnetosheath used extensively in Earth studies [e.g. *Longmore et al.*, 2006; *Cooling et al.*, 2001; *Petrinec*, 2012] including Mercury and Saturn. The model describes a static magnetic field in the magnetosheath by means of a scalar potential which is a solution to Laplace's equation. The model uses the IMF as input, imposes jump conditions at the first boundary (bow shock) and a boundary condition is set such that the magnetic field is zero at the second boundary (magnetopause). Two key features of this model are that the boundaries are paraboloids symmetric along the planet-Sun line and field-flow coupling effects are not taken into account.

A formal procedure to carry out a comparison of the observed magnetosheath magnetic field with the model and hence testing the model's ability to predict the draping pattern is best done via a statistical study (such as in *Longmore et al.* [2006]). This work used a multi-spacecraft technique with the ACE spacecraft providing the IMF data while the Cluster spacecraft provided good coverage of the magnetosheath. The effects of boundary motion were overcome by normalizing each measurement within the magnetosheath to a local position between the two boundaries. As for the effects of solar wind variability, the magnetic field measurements in the magnetosheath were normalized to corresponding time-lagged ACE measurements. Since the standoff distance of a boundary is determined only at the time of crossing, it is not possible to determine the locations of both boundaries as well as the upstream conditions simultaneously at any given time using a single spacecraft such as we have with Cassini at Saturn. In order to mitigate transient effects, a case study approach is taken using four time series when the IMF orientation is relatively steady throughout Cassini's inbound traversals in the magnetosheath. The model is employed to predict the draping pattern of the field lines in the magnetosheath using the IMF as measured by Cassini before crossing the bow shock. The clock angles of the predicted field lines will then be compared with observations by Cassini during its magnetosheath traversals. An instructive comparison would be of the magnetic field against maps of the near-magnetopause region by *Desroche et al.* [2013] and/or profiles showing how the magnetic field evolves as a consequence of non-axisymmetry along the subsolar line in the magnetosheasth [*Erkaev et al.*, 1996; *Farrugia et al.*, 1998]. The aforementioned limitations, however, do not allow for the variability to be separated from being spatial or temporal even for the steadiest traversals. The standard deviation of the magnetic field along the magnetosheath was found to be up to 15° for steady cases and thus cannot distinguish a near-magnetopause magnetic field orientation from an adjacent one.

In this section we conduct a case study to examine the correspondence between the observed draping in the magnetosheath with that predicted by KF94 under different IMF orientations. By testing its validity, this will also provide additional means in approximating the magnetic field in a region where Cassini is not present to make observations. The magnetosheath during these four traversals was measured to have relatively steady IMF orientations throughout the traversal, also used in the work by *Masters et al.* [2014]. This one-to-one approach is not suitable with the MHD simulation since the grid size near the magnetopause and bow shock is ~0.5 $R_s$. Typically a traversal can span six grids and this would thereby return six vectors which is not useful for this study considering the high variability of the magnetic field in the magnetosheath. The inset in Figure 6c shows the four different clock angles considered. They are categorized into a higher $B_z$ component, labelled S and N for southward and northward orientations respectively, and a lower $B_z$ component labelled O1 and O2. Figures 6a and 6b show the pairs of observed (red) and KF94 predicted (black) 3D vector plots for the S and O1 orientations during their traversals respectively.

These traversals started and ended at ~5% of the transit time from each boundary to alleviate the effects of near-boundary activities (such as the PDL near the magnetopause) which are not accommodated by KF94. The four points corresponding to every time history do not represent the same fractional distance since each traversal had a different transit time. Since boundaries can only be observed during the time of the crossing, determining an accurate fractional distance in the magnetosheath during a traversal is not possible. However, full traversals such as these in the case study are generally caused by magnetospheric compression/expansion and we therefore expect to traverse the magnetosheath monotonically from bow shock to magnetopause or vice versa. IMF orientations with higher $B_z$ (S and N) show a significant clock angular rotation between the observations and KF94. This rotation becomes less pronounced with increasing equatorial orientation (O2 then O1). In addition, the rotation tends to increase nearer the magnetopause for S, N and O2 and there is no correlation for O1.

## 4. Discussion

Our results present the dayside draping of the observed magnetosheath, which were mainly originated as the Parker spiral configuration. We assessed the asymmetry of the field line curvatures using MHD simulations with the expectation that the asymmetry of the magnetopause will be manifested. Our analysis from observations were limited due to the lack of an upstream monitor to mitigate the effects of solar wind variability and boundary motion, both of which dictate Cassini's fractional position in the magnetosheath as well as restricting the model comparisons to (quasi-) steady state traversals.

The MHD simulations show a clear-cut asymmetry in the angular variations with longitude and latitude for IMF vectors aligned in the equatorial and meridional planes respectively. This revealed ~15% more curvature from the angular gradients calculated in the meridional over the equatorial planes and this is indicative of the effect of the poles being more flattened than the flanks on the global magnetic structure. The MHD simulation is in broad agreement with measurements of the azimuthal angle of the magnetosheath across a wide range of longitudes by having a similar gradient to the distribution of observed azimuthal angles on the equatorial plane. The statistics for the meridional angle, on the other hand, are weak due to the lack of high latitude coverage as well as the very few incidents of steady northward or southward IMF orientations. The five events studied exhibit some trend compared to the MHD simulated sets although they are not statistically conclusive and there is some spread owing to their non-perfect alignment as well as some variability in the magnetosheath. See inset in Figure 6c for S, N and O2 (three of the five events).

Nevertheless, despite the magnetic field lines being organised equatorially, the scattering of the observations in Figures 5a and 5b reveal the significance of non-zero meridional angles. This is also highlighted in Figure 2. The distributions of the meridional angles in both flanks show substantially more non-zero meridional angles compared with the Parker spiral upstream of Saturn [*Jackman et al.*, 2008] which generally has a small $B_z$ component. Apart from the temporal variability of the solar wind, it is likely that the non-axisymmetry of Saturn's magnetosphere is responsible for twisting the magnetic field out of the equatorial plane; consistent with the prediction by *Erkaev et al.* [1996] and *Farrugia et al.* [1998].

Using KF94, we see better correspondence between the clock angles of the predicted and observed magnetic field vectors with a lower $B_z$ component. *Longmore et al.* [2006] conclude this is due to field-flow coupling effects in which the bulk flow acts on the magnetic field lines to twist the IMF. These contribute towards rotating the magnetic field lines in the direction of the accelerated flankward flow. Consistent with the study, a higher $B_z$ component leads to a rotation in the clock angle relative to KF94. However with limited coverage of the magnetosheath and plasma instruments, the role this plays in Saturn's magnetosheath cannot be corroborated. Here in particular, we find a more significant rotation for the higher $B_z$ compared to Earth and this could also possibly be due to the significant meridional confinement at Saturn. Thus the negligence of both field-flow coupling effects and boundary asymmetry lead to departures from the prediction and this is indicative of their role in the overall draping pattern. Nevertheless, KF94 has a potential reliability in determining the draping pattern and ultimately the shear angle near the magnetopause when Cassini is in the IMF region given that the orientation is near equatorial and assuming it will remain fairly steady. The Michigan Solar Wind Model (MSWiM) is commonly used to predict conditions of the solar wind upstream of Saturn. This is a one-dimensional (1-D) magnetohydrodynamic code that uses near-Earth spacecraft measurements as boundary conditions at 1 AU and simulates the evolution of solar wind parameters along heliocentric distances through to 10 AU (beyond Saturn's orbit). Comparisons of the predicted and observed data are found to be most agreeable when the solar wind exhibited a recurrent pattern during the declining phase of the solar cycle. Furthermore, the accuracy of speed propagations was most accurate within approximately ±75 days of apparent opposition (Sun-Earth-Saturn alignment taking into account transit time of the solar wind between Earth and Saturn) [*Zieger and Hansen*, 2008]. KF94 can be used in conjunction with propagations from MSWiM and remote auroral observations to investigate reconnection at the magnetopause when Cassini is in the IMF or magnetosphere during the aforementioned apparent opposition.

Future work will focus on the effect of the non-axisymmetry on the evolution of and twisting of the magnetic field lines out of the equatorial plane. It will be interesting to compare with works by *Desroche et al.*, [2013]. More observations and specifically steady traversals will be investigated which could potentially support these findings.

Acknowledgments

AHS acknowledges useful discussions with S. J. Schwartz and M. O. Archer as well as the support of the Royal Astronomical Society. We acknowledge the support of MAG data processing/distribution staff. This work was supported by UK STFC through the award of studentships (AHS) and research grants to Imperial College London. AM acknowledges the support of the JAXA International Top Young Fellowship Program. XJ acknowledges the support by NASA through grant NNX12AK34G and by the Cassini mission under contract JPL 1409449.

SULAIMAN ET AL.: MAGNETIC STRUCTURE OF SATURN'S MAGNETOSHEATH

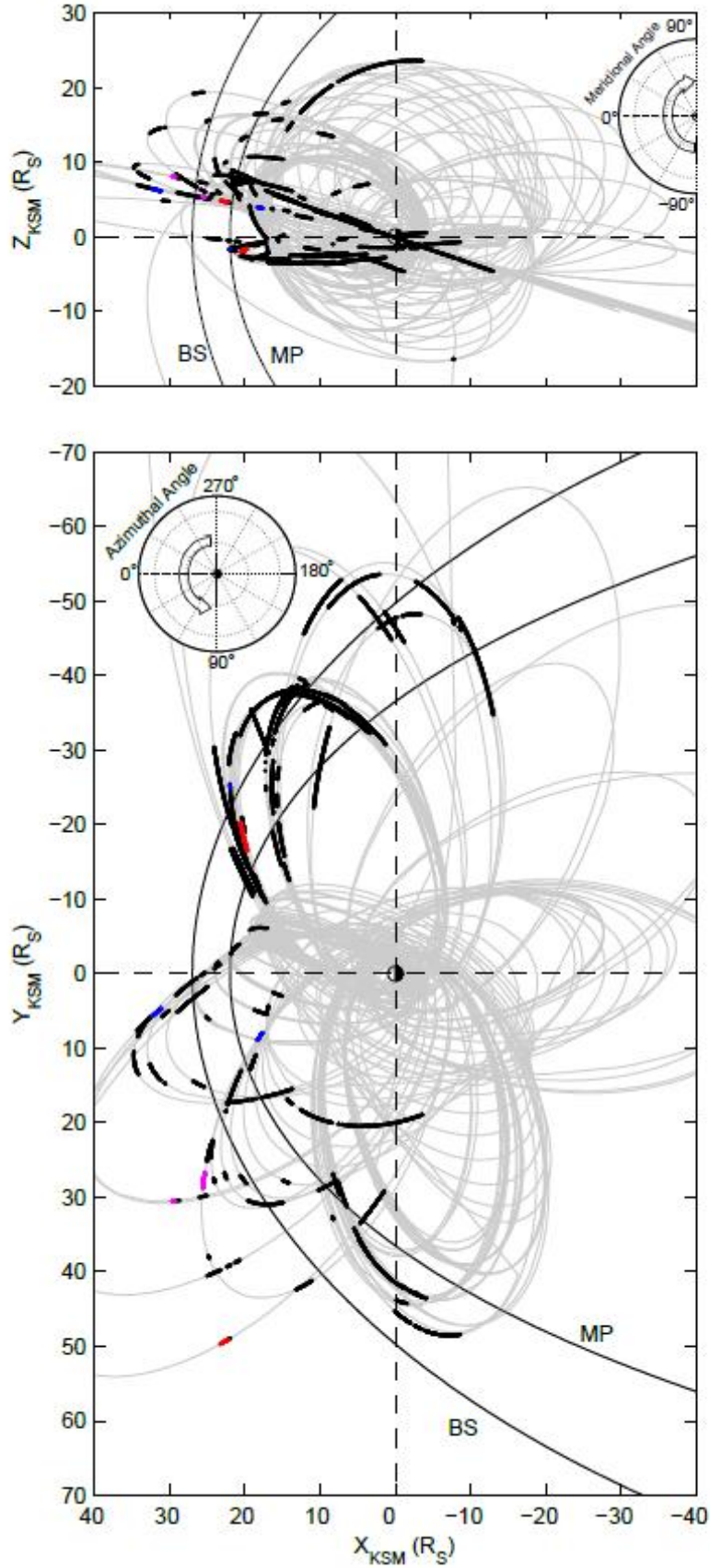

**Figure 1**: An overview of the trajectory of the Cassini spacecraft (gray) between July 2004 – December 2010 with observed magnetosheath boundary-to-boundary traversals (black) indicated and projected onto (a) the *X-Z* and (b) the *X-Y* planes. The IMF orientation was relatively steady throughout the traversals in colour. Blue indicates northward IMF, red indicates traversals used in the case study and magenta indicates a combination of both. In both figures the projections of the *Kanani et al.* [2010] magnetopause and *Went et al.* [2011] bow shock models are shown with median subsolar distances of 22 $R_S$ and 27 $R_S$ respectively for a solar wind dynamic pressure of ~0.02 nPa. The medians of the sets of respective boundary crossings are used so that errors in the models or extreme events which produce anomalous estimates do not significantly skew the determination of the typical subsolar distance. The inset on the top and bottom panels define the meridional and azimuthal angles respectively.

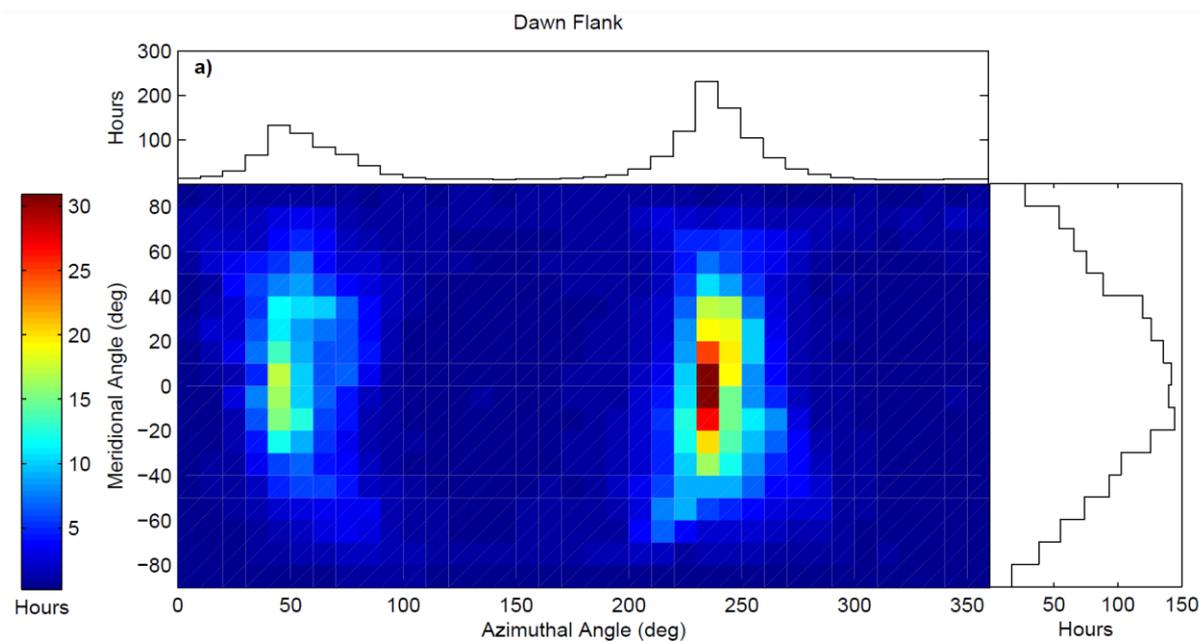

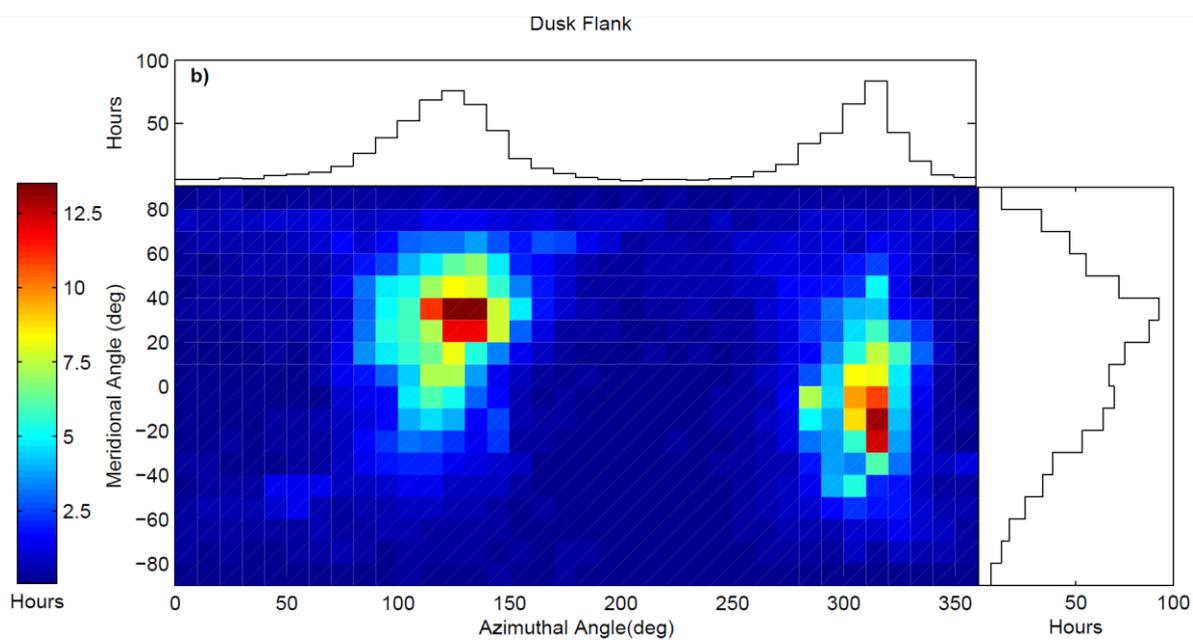

**Figure 2**: The statistical configuration of the magnetic field on the a) dawn flank and b) dusk flank of the magnetosheath. The 2-D histograms are color scaled to the length of time the magnetic field has been observed in a particular combination of meridional and azimuthal directions. The adjacent histograms project these angles individually.

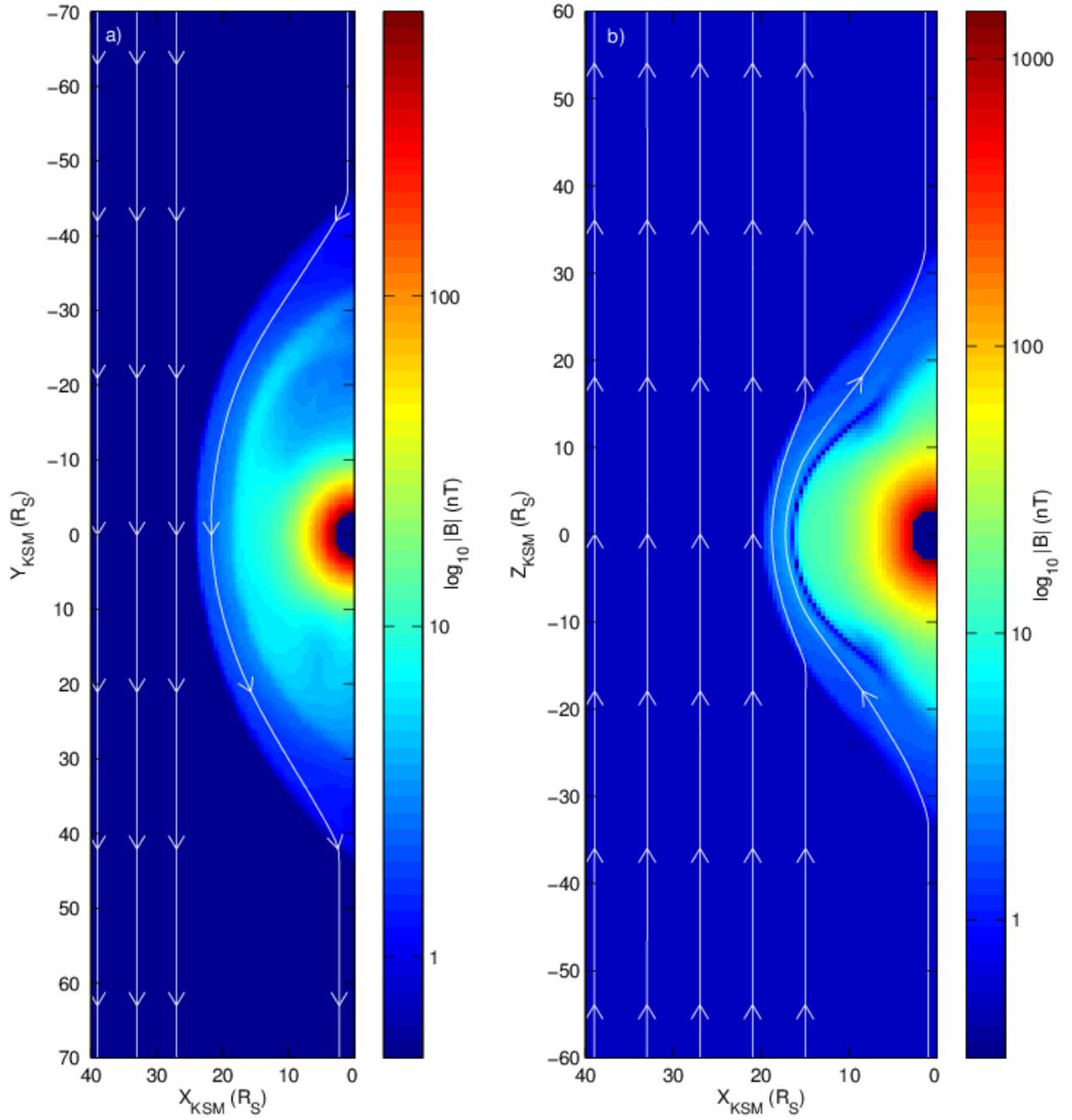

**Figure 3**: MHD simulation snapshots of dayside draping of the magnetic field for a) duskward and b) northward IMF orientations along the *X-Y* ($Z_{KSM} = 0$) and *X-Z* ($Y_{KSM} = 0$) planes respectively. The color scales represent the logarithmic magnetic field magnitude and the white arrows are magnetic field lines which bend upon encountering the bow shock (first anti-planetward boundary) and arrange tangentially to the magnetopause (boundary enclosing high magnetic field magnitude region).

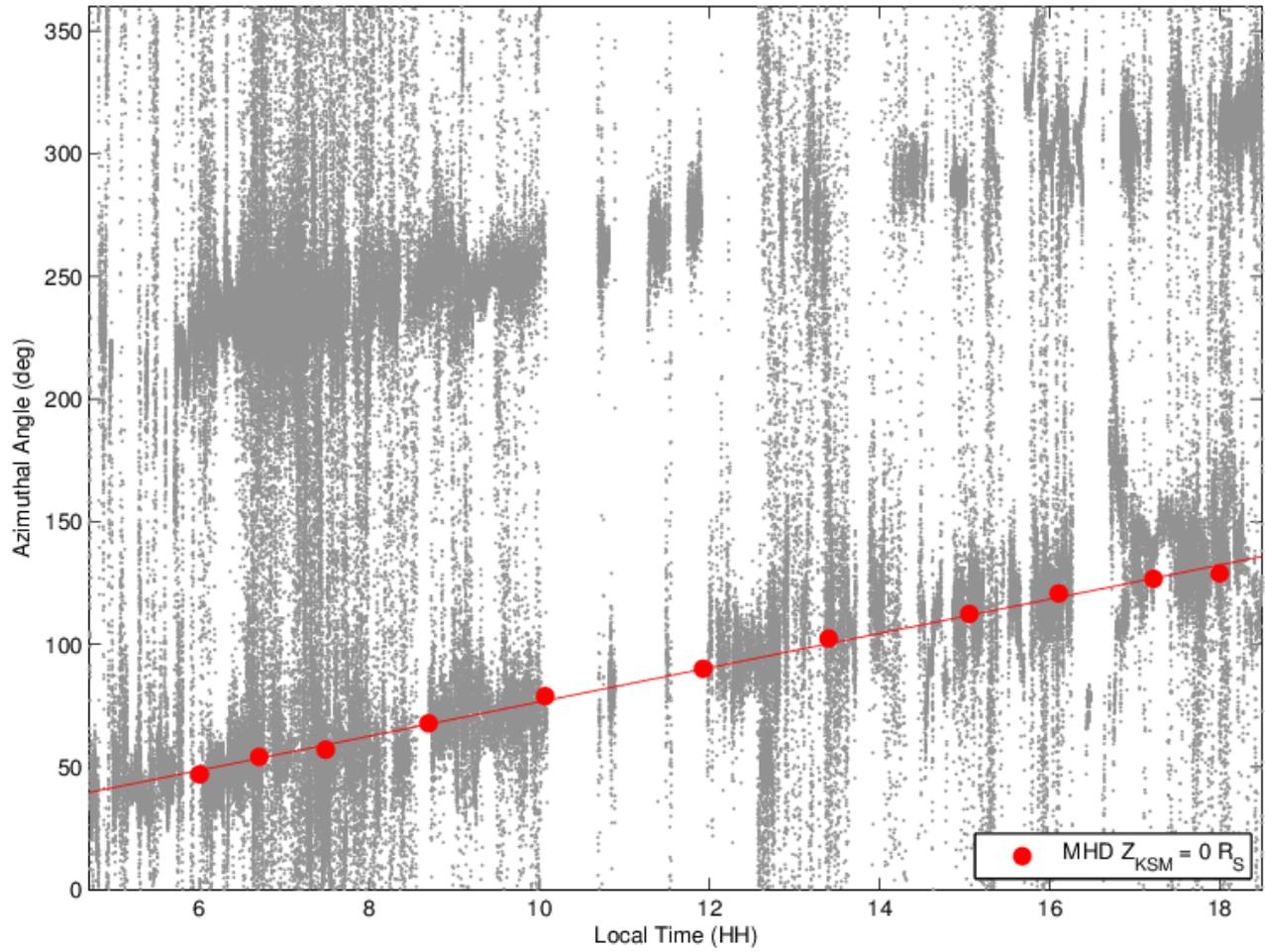

**Figure 4**: Distribution of the observed azimuthal angle in the magnetosheath with local time projected on a plane. Overlain are the MHD simulated angles at different local times for a duskward IMF.

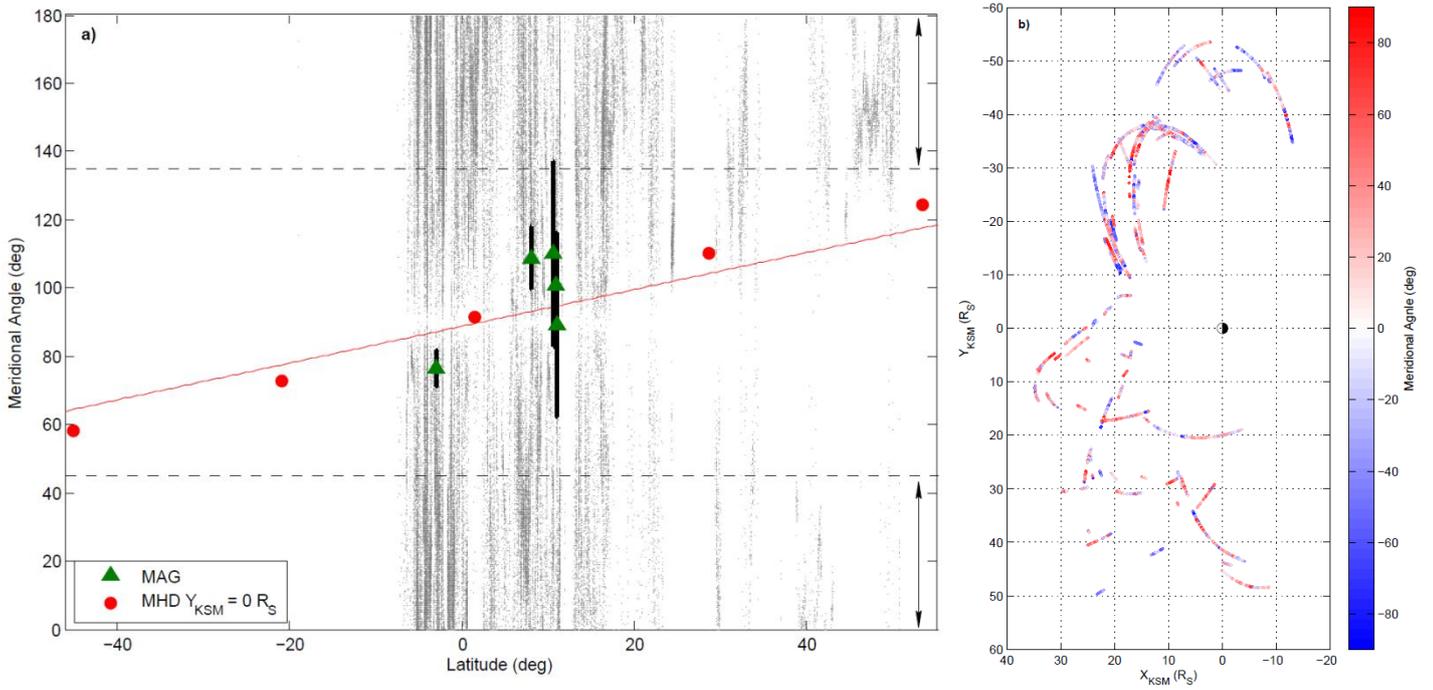

**Figure 5**: a) Distribution of the observed meridional angle for northward orientations in the magnetosheath with latitude projected on a plane. The range is extended to 180° to distinguish between +X and –X. The double-headed arrows indicate regions within ±45° of the equator. Overlain are the mean observed angles at five traversals with error bars when the IMF was within ±45° of north and steady (green) and MHD simulated angles at different latitudes for a northward IMF (red). b) Distribution of observed meridional angle on the equatorial plane.

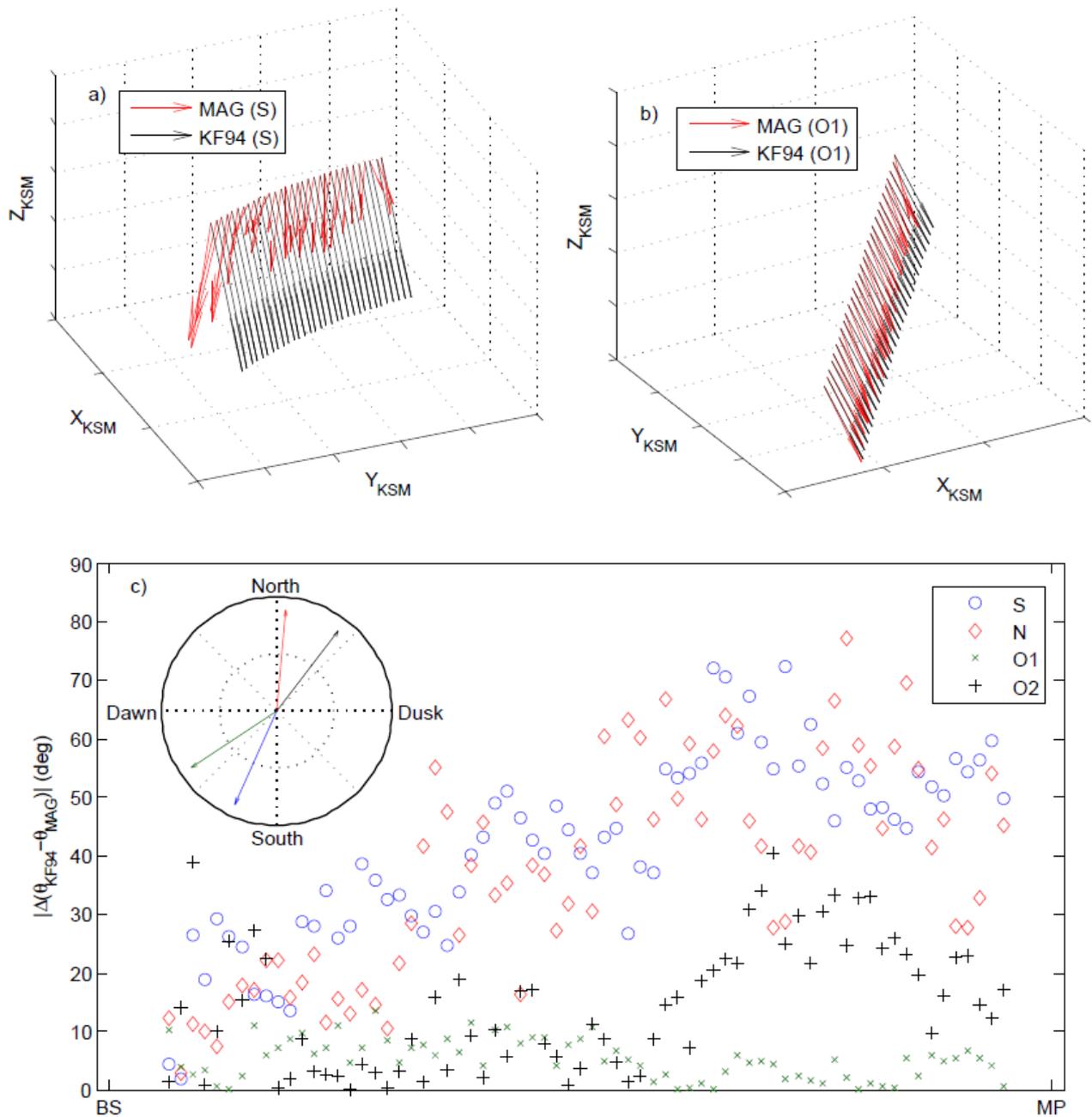

**Figure 6**: Observed and KF94 predicted vector plots of draped field lines throughout magnetosheath traversal for a) near-southward and b) near-equatorial orientations. c) The angular difference between observed and KF94 for different IMF clock angles (inset).

**Table 1:** Upstream conditions of MHD simulations

| Direction | $B_x$ (nT) | $B_Y$ (nT) | $B_Z$ (nT) | $D_p$ (nPa) | Grid Resolution near dayside magnetosheath ($R_s$) |
|---|---|---|---|---|---|
| Northward | -2.41E-05 | -1.28E-05 | 4.96E-01 | 4.22E-04 | 0.5 |
| Duskward | 3.38E-05 | 4.94E-01 | 3.26E-05 | 2.43E-04 | 0.5 |